# PentestChain: A Cost-Aware, MCP-Orchestrated Framework for Automated Penetration Testing with Free-Tier LLMs

Rushabh Vipulkumar Patel
Computer Security Laboratory, University of Wales Trinity Saint David
London, United Kingdom
2436066@student.uwtsd.ac.uk

Dipo Dunsin
Computer Security Laboratory, University of Wales Trinity Saint David
London, United Kingdom
d.dunsin@uwtsd.ac.uk

Mohammed Almaiah
King Abdullah IT School, The University of Jordan
Amman, Jordan
m.almaiah@ju.edu.jo

Mohamed Chahine Ghanem*
School of Computer Science and Mathematics, Keele University
Newcastle-under-Lyme, United Kingdom
m.ghanem@keele.ac.uk

## Abstract

AI-driven penetration testing has been demonstrated with premium frontier models such as GPT-4, but the per-engagement token cost makes continuous, automated testing unaffordable for the smaller organisations that need it most. This paper presents PentestChain, a ten-phase automated penetration testing framework that couples a curated, deterministic exploit map with a cost-aware AI cascade—a local Ollama model (`qwen2.5:7b`) first, then free-tier OpenRouter and Cerebras, with a rule-based fallback that always produces output—and exposes the full pipeline through a Model Context Protocol (MCP) server with eleven tools. We make three contributions. First, we treat U.S.-dollar cost per engagement as a measured, first-class evaluation metric and show that a 7B-parameter local model, kept off the critical path by a deterministic backbone, sustains end-to-end operation at zero measured paid-API cost. Second, we analyse the attack surface that an MCP-exposed offensive engine introduces, grounding a four-position threat model in the 2025 MCP incident record (the CVE-2025-6514 remote-code-execution flaw in `mcp-remote`, the `postmark-mcp` supply-chain backdoor, and the tool-poisoning/rug-pull/line-jumping class), and contribute four mitigations. Third, we specify a reproducible, containerised evaluation protocol aligned with the standardised testbeds now expected at top-tier venues—AutoPenBench, a Cybench subset, and the PentestGPT 182-sub-task benchmark—with multi-trial statistics ($\geq$ 10 trials per configuration, `pass@`$k$, non-parametric significance tests and effect sizes) and direct, same-testbed reproduction of the PentestGPT and PentestAgent baselines rather than citation of their published numbers. On the legacy targets measured to date, the framework detected 26 services, enriched 34 CVEs, produced 11.5 ± 0.6 ready exploits with confirmed shell and Meterpreter sessions, and generated remediation for every detected service, at zero measured paid-API cost, with a mean pipeline time of 21.7 ± 4.4 minutes and a 3.6% fallback rate to the deterministic path.

*Corresponding author.

## 1 Introduction

New-CVE volume rose roughly 32% year on year in 2024, exceeding forty thousand newly published records [12, 47], and the average data-breach cost reached USD 4.88 million [37]; the Verizon 2024 breach study likewise reports a sharp rise in exploitation of vulnerabilities as an initial access vector [64]. Yet the ISC2 2024 study estimates a global cybersecurity workforce gap of 4.8 million practitioners [39]. Penetration testing catches these problems before an attacker does—but at a cost and cadence that gap makes hard to sustain. The market response, penetration-testing-as-a-service, is projected to grow past USD three billion by the end of the decade [43], evidence of demand that human-delivered testing alone cannot meet at the frequency modern change velocity requires.

A wave of AI-driven tools has appeared in response, from PentestGPT [14] and AutoPentester [27] to remediation-aware and MCP/multi-agent systems [4, 20, 36, 57, 65] (Section 2). Almost all share one assumption: unrestricted access to a premium frontier model such as GPT-4, whose token cost makes continuous testing unaffordable for the organisations that need it most. The dependence is not merely economic but methodological: surveys of offensive-security benchmarking find only about a quarter of published prototypes evaluate a local or small model at all [30], leaving the free-tier regime largely uncharacterised.

This paper asks: *can a structured, pipeline-based framework driven by free-tier and locally hosted LLMs, backed by a deterministic fallback, achieve effectiveness comparable to premium-model frameworks while reducing per-engagement cost to zero?* The economic stakes are concrete. Fang et al. report that an autonomous GPT-4 agent exploits 87% of a set of one-day vulnerabilities at an average of USD 3.52 per run—roughly 2.8× cheaper than human labour, but still a recurring per-engagement charge that compounds under continuous testing [21]. Driving that charge to zero, rather than merely lowering it, is what makes exploitation-grade testing available to organisations that cannot budget frontier-model tokens at fleet scale.

To answer this, we present PentestChain, a ten-phase automated framework running inside a Kali Linux Docker container with a PostgreSQL companion. Its key design decisions are: (1) a curated rule-based exploit map that carries the offensive logic deterministically, confining the LLM to nine bounded roles; (2) a cost-aware AI cascade that tries a local Ollama model first, then free-tier cloud providers, then a rule-based fallback; and (3) exposure of the full pipeline through an MCP server with eleven tools, so any compliant host can drive the engine. These choices make a 7B-parameter local model sufficient, because the model is never on the critical path for exploitation.

The contributions of this paper are:

- A free-tier-first, cost-aware architecture for AI-driven penetration testing in which a deterministic backbone makes a modest LLM sufficient.
- The first cost model for AI-assisted penetration testing that treats dollar cost per engagement as a measured, first-class metric.
- A four-position threat model and four mitigations for exposing offensive tooling through MCP, grounded in the 2025 MCP vulnerability record.
- A reproducible, containerised evaluation protocol aligned with standardised testbeds (AutoPenBench, a Cybench subset, and the PentestGPT benchmark), specifying multi-trial statistics and same-testbed reproduction of the PentestGPT and PentestAgent baselines.
- Empirical results across engagements on five heterogeneous targets (legacy Linux, Windows, web-focused, containerised, and modern Linux) showing comparable effectiveness to premium-model frameworks at zero measured cost.

## 2 Related Work

### 2.1 Automated Penetration Testing

Automated penetration testing spans a spectrum from reinforcement learning to LLM-driven agents. DeepExploit [58] trained an A3C agent to select Metasploit modules from banners but did not generalise beyond its training targets. Schwartz and Kurniawati [55] formalised penetration testing as a POMDP, showing model-free RL could find optimal attack paths in simulation. Microsoft's CyberBattleSim [45] provided an abstract network-attack environment that made this line reproducible, and Erdődi and Sommervoll's Agent Web Model [18] extended the RL formulation to web hacking. Ghanem et al. [25] demonstrated deep RL exploitation against Metasploitable-style targets, though training cost and brittleness to banner variation remained. The simulation-to-reality gap is the recurring weakness of the RL approach: an agent optimised in an abstract or simulated environment rarely transfers to the exact banner strings and tool-output idiosyncrasies of a live host [61].

PentestChain sidesteps this by pairing a curated exploit map with an AI fallback used only when no rule matches, trading end-to-end learning for deterministic, explainable exploit selection that needs no training data.

### 2.2 LLMs in Offensive Security

The use of LLMs for offensive security was crystallised by Happe and Cito's early demonstration that a conversational model could drive privilege-escalation attacks under human guidance [32]; their subsequent 2025 survey [31] finds LLMs are strong at information retrieval and output interpretation but weak at long-horizon planning and reliable command generation. PentestGPT [14] showed LLMs hold enough knowledge for the task but required a human to copy each command. AutoPentester [27] closed that gap with direct tool calling. PenHeal [36] added remediation, improving vulnerability coverage by 31%. PentestAgent [57], the closest methodological precedent to our design, decomposes an engagement into intelligence gathering, vulnerability analysis and exploitation stages driven by collaborating agents with RAG, and reports stage-level completion rather than a single figure—the reporting style we adopt for our own ablation. D-CIPHER [26, 62] pushes the multi-agent direction further with planning and heterogeneous execution roles. The Fang et al. line of work [21–23] established that frontier agents can autonomously hack websites, exploit one-day CVEs given their description, and, in teams, chain toward zero-days.

A recurring finding across this literature is a *knowledge–execution gap*: small and mid-sized models frequently produce a correct natural-language description of an attack (for example, identifying an IDOR or drafting a valid SQL-injection payload) yet fail to *carry it out* through tool calls, and typed tool interfaces alone have been shown to recover a large fraction of that lost performance [30]. This directly motivates PentestChain's design: the deterministic exploit map supplies the execution that a 7B model cannot reliably generate, so the model is asked only for the labelling and ranking it does well.

### 2.3 Benchmarks for Security Agents

The field has converged on a small set of standardised, containerised testbeds, and top-tier reviewers increasingly expect evaluation on them rather than on bespoke targets. Cybench [66] assembles 40 professional CTF tasks from HackTheBox, SekaiCTF, Glacier and HKCert, using First Solve Time as a difficulty proxy and offering both unguided and subtask-guided modes. AutoPenBench [28] contributes 33 milestone-scored tasks (22 in-vitro plus 11 real-world CVE) in Docker, with reported autonomous success far below human-assisted success, and has become a common reference testbed. The PentestGPT benchmark [14] itself—13 targets decomposed into 182 sub-tasks spanning the OWASP Top 10—remains the standard instrument for granular, sub-task-level scoring. NYU CTF Bench [56] adds a scalable, open-source CTF dataset built specifically for evaluating LLMs on offensive-security tasks, broadening the pool of standardised instruments. Happe and Cito's review of benchmarking practice [30] catalogues the metrics and testbeds in use and documents the small-model evaluation gap noted above. Section 6 adopts these instruments directly.

## 2.4 Cost and Accessibility

Published frameworks evaluate almost entirely on effectiveness. PentestGPT and AutoPentester use GPT-4 [14, 27]. The Ezetta–Feng toolkit reports its headline results with GPT-4.1, and the same-named multi-agent preprint reports, with GPT-4.1, success rates of 87.3% on information gathering, 62.3% on vulnerability discovery and 56.6% on exploitation across more than one hundred VulHub and NVD vulnerabilities [4, 20]. PentAGI accepts local models but assumes a premium frontier model for serious work [1, 65]. None report a dollar cost per engagement or ask whether a free-tier model would suffice.

If effectiveness depends on a premium model, exploitation-grade testing is available only to organisations that can pay for frontier-model tokens at scale—excluding exactly the under-resourced organisations the workforce-gap literature identifies as most exposed [39]. Treating per-engagement cost as a first-class metric is thus both novel and practically relevant, and because the deterministic backbone removes the need for a frontier model, the cost reduction is structural rather than a quality-for-price trade. This is the gap PentestChain targets.

## 2.5 MCP: From Protocol to Attack Surface

The Model Context Protocol [6, 7], released by Anthropic in November 2024, defines a JSON-RPC architecture for tool calls between AI hosts and servers. Its adoption has been rapid and its security is now an active research area, with a documented 2025 incident record [13, 38, 40, 60] and emerging taxonomies [34, 67]. Dedicated security benchmarks have begun to appear—MCPSecBench [3] provides a systematic playground for probing MCP implementations—and industry analyses now catalogue the classes of flaw practitioners must address [19]. A parallel strand studies prompt injection in tool-integrated agents and the protocol's exposure to it [44]. Because PentestChain exposes offensive capability over MCP, that surface shapes its design; we defer the full treatment to the threat model in Section 3.

## 2.6 Positioning Against MCP-Based Pentest Systems

Two distinct 2025 works share the name "PentestMCP." The Ezetta–Feng *toolkit* [20] is a library of MCP server implementations spanning scanning, enumeration, fingerprinting, vulnerability scanning, exploitation and post-exploitation—a set of building blocks rather than an orchestrated, cost-instrumented pipeline. The multi-agent *framework* [4] couples LLMs, MCP and RAG behind a task graph and is evaluated on 100+ VulHub/NVD CVEs with GPT-4.1. PentestChain differs from both on the axis this paper measures: it places the offensive logic in a deterministic map so that a *free-tier or local* model suffices, reports *measured* zero dollar cost, and analyses the MCP surface it exposes rather than only consuming MCP as plumbing. Relative to PentAGI [65] (premium-first, local-capable) and to fine-tuned local agents such as xOffense [5], PentestChain requires no training and keeps a deterministic correctness floor beneath the model. We reproduce PentestGPT and PentestAgent as same-testbed baselines in Section 6 rather than comparing against their published numbers.

# 3 Threat Model

Exposing an autonomous exploitation engine through a general-purpose tool-calling protocol creates an attack surface distinct from that of the target network. This section defines that surface. Throughout, the *engine* is the PentestChain Flask/MCP process, the *host* is the MCP-compliant client that drives it (for example, a desktop LLM application), and the *target* is the system under test. Figure 1 summarises the resulting attack surface and the mitigations that address it.

## 3.1 Assets and Adversaries

The assets to protect are (i) the exploitation capability itself (unauthorised use of the engine against a system the operator does not own), (ii) the integrity of the engine's decisions (which host instructions it honours), and (iii) the confidentiality of engagement data (scan results, credentials recovered during post-exploitation, target inventory). We consider four adversary positions, aligned with the MCP incident record.

**A1: Malicious client.** A host directs the engine at a target the operator is not authorised to test, turning a defensive tool into an unauthorised-access instrument.

**A2: Compromised or hostile server/tool poisoning.** A malicious MCP server—or a poisoned tool description on an otherwise trusted server—injects instructions the host's model then acts upon. Invariant Labs demonstrated that such instructions can be hidden in the tool *metadata* delivered in the `tools/list` response *before any tool is invoked*; Trail of Bits terms this "line jumping." A "rug pull" silently redefines an approved tool after first use [38, 60]. The `postmark-mcp` package is the canonical real-world instance: a clean mirror of an official server through v1.0.15 that, in v1.0.16, added a one-line BCC backdoor exfiltrating all outgoing mail—the first malicious MCP server observed in the wild [13].

**A3: Client remote code execution.** A flaw in the client's protocol handling lets a hostile server execute code on the operator's machine. CVE-2025-6514 (CVSS 9.6) is the concrete case: a crafted `authorization_endpoint` returned during the OAuth flow was passed to the OS URL handler by `mcp-remote` (versions 0.0.5–0.1.15), yielding arbitrary command execution and full client compromise [40]. This is an attack *on the operator through the protocol*, independent of the target network.

**A4: Man-in-the-middle / confused deputy.** A network adversary replays or reorders tool calls, or induces the engine to use its legitimate target authorisation on the attacker's behalf.

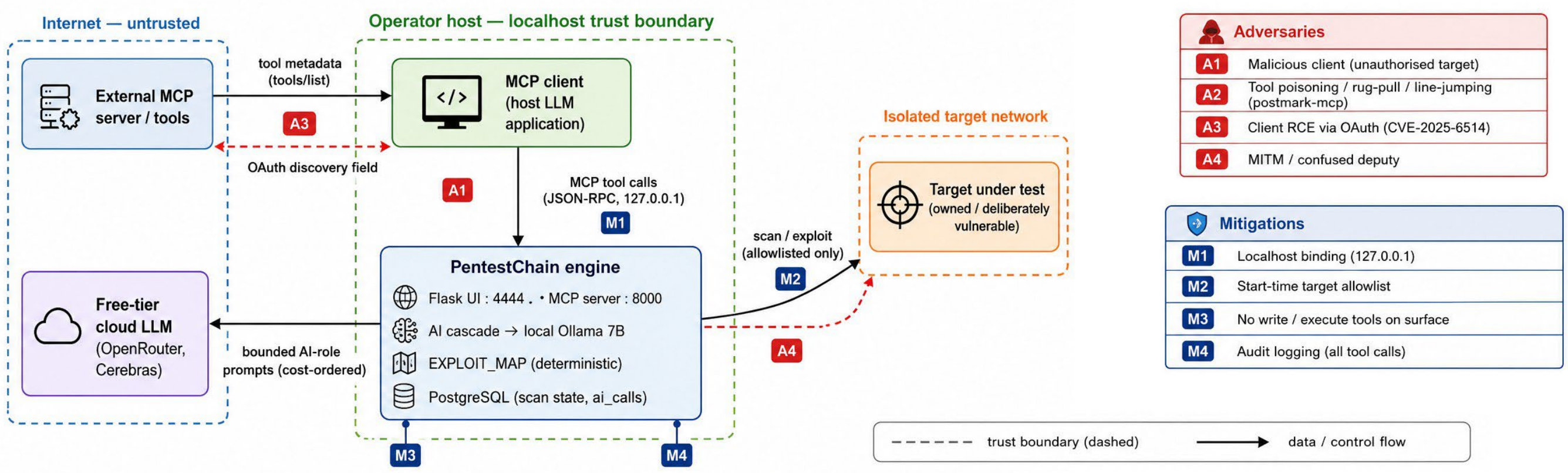

**Figure 1: PentestChain's MCP attack surface and mitigations.**

## 3.2 Specification-Level Considerations

MCP carries JSON-RPC messages; its remote HTTP transport now mandates OAuth 2.1 with PKCE, using Protected Resource Metadata (RFC 9728) and Resource Indicators (RFC 8707)—the mechanism that binds a token to one server and prevents pass-through between servers [10, 41, 46]. Two facts matter here. First, only a fraction of deployed servers implement the mandated authorisation, so the specified protections are frequently absent in practice [34]. Second, A3 shows the OAuth flow is itself part of the attack surface—a discovery field returned during authorisation was the injection vector—so "use OAuth" is necessary but not sufficient.

## 3.3 Mitigations

PentestChain answers these four positions with four controls, each mapped to the adversary it addresses.

(1) *Localhost binding* (A1, A4): the engine binds to `127.0.0.1` only, removing the cross-host network surface and the replay vector that a remote client would need.
(2) *Target allowlisting* (A1): every scan tool validates the target IP against an allowlist fixed at server start, so a malicious host cannot redirect the engine to an unauthorised system even if it reaches the tool interface.
(3) *Excluded writes* (A2): file-system writes and remediation-script execution are excluded from the tool surface, so a poisoned tool description or rug-pulled tool cannot induce the engine to persist or execute attacker-controlled changes on the host.
(4) *Audit logging* (A2, A3, A4): every tool call is logged with timestamp, tool name and arguments, giving the tamper-evident record needed to detect line-jumping, rug-pulls and replayed calls after the fact.

These controls are defence-in-depth, not a completeness claim: A3 in particular is a client-side flaw that the engine cannot fully close, which is why we recommend pinning client versions and treating tool descriptions as untrusted input. We revisit the residual risk in Section 9. Table 1 summarises the mapping from adversary to control and the residual exposure that remains.

**Table 1: Adversary → vector → control mapping, with residual risk**

| Adv. | Vector | Control(s) | Residual |
|---|---|---|---|
| A1 | Unauthorised target via host | Localhost bind; target allowlist | Low |
| A2 | Tool poisoning / rug-pull / line-jumping | Excluded writes; audit log | Medium |
| A3 | Client RCE (e.g. CVE-2025-6514) | Audit log; pin client (external) | High |
| A4 | MITM / confused deputy | Localhost bind; audit log | Low |

Residual risk is the exposure remaining after the default controls; A3 is client-side and cannot be closed by the engine alone.

# 4 System Architecture

## 4.1 Four-Layer Design

PentestChain has four layers (Fig. 2). The presentation layer is a Flask application serving the web UI on port 4444 with real-time events via Flask-SocketIO. The orchestration layer is the `run_full_scan` state machine with per-phase logging. The execution layer wraps Nmap, the Metasploit Framework (via pymetasploit3), Nikto, sqlmap and WPScan as subprocess calls with structured output parsing. The persistence layer is PostgreSQL, holding scan state across container restarts. Two cross-cutting services sit alongside: the AI helper module (invoked from nine bounded points) and the FastMCP server on port 8000.

## 4.2 Cost-Aware AI Cascade

The cascade is the heart of the free-tier thesis. Providers are tried in strict cost order: (1) a local Ollama model [49] (`qwen2.5:7b` [52], zero marginal cost, data stays on host); (2) free-tier OpenRouter; (3) free-tier Cerebras; and (4) a deterministic rule-based result that always succeeds. The `call_ai_fast` function moves to the next provider on any non-200 response, 429 rate limit or timeout, and marks a provider as dead for 90 seconds. Per-minute guards stay within free-tier ceilings. Algorithm 1 states this precisely. The invariant that makes the free-tier claim defensible is that the loop *cannot fail*: the deterministic branch $D(\cdot)$ is total, so every call site

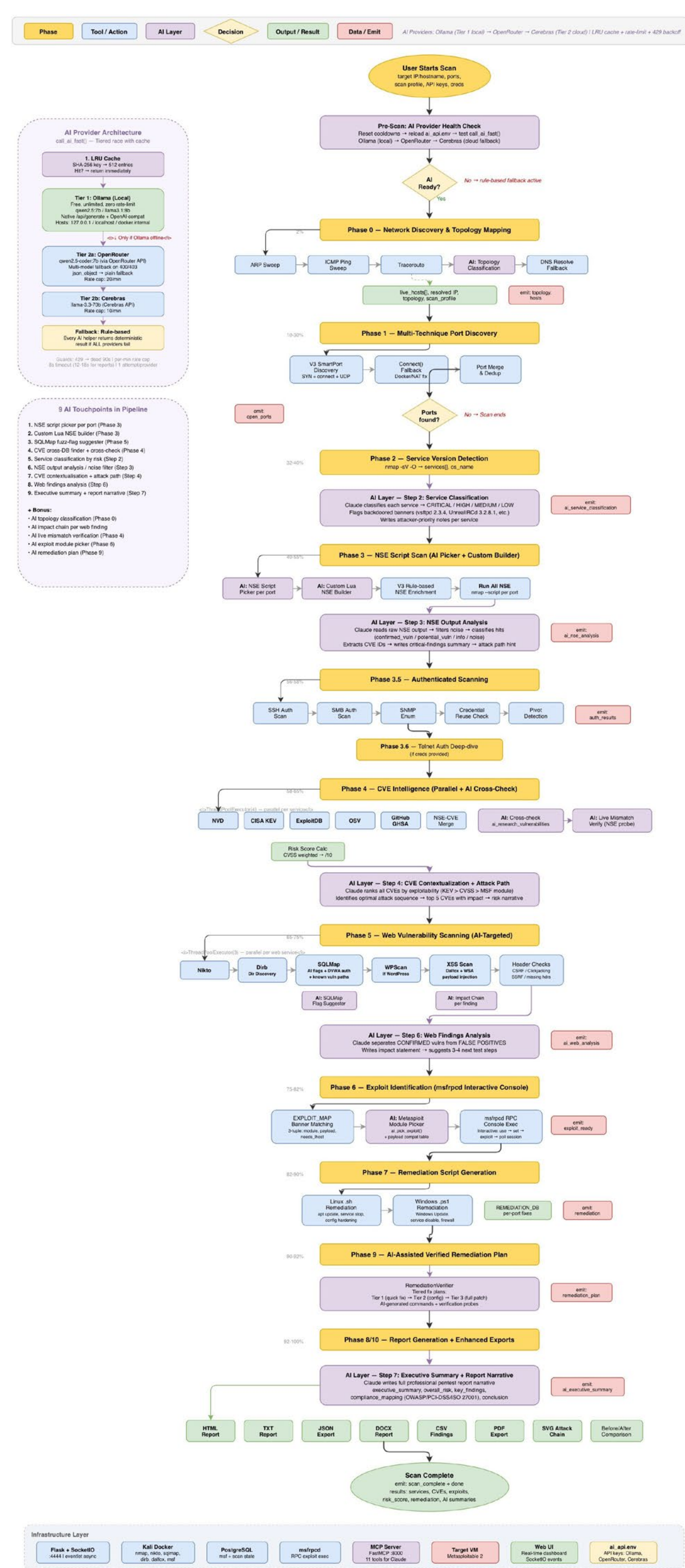

Figure 2: PentestChain pipeline workflow.

**Algorithm 1** Cost-aware cascade `call_ai_fast`

**Require:** prompt $p$, role $r$; provider list $P = \langle$Ollama, OpenRouter, Cerebras$\rangle$ in cost order; deterministic fallback $D_r$
**Ensure:** a valid result for role $r$ (never $\perp$)
1: **for** $i \leftarrow 1$ to $|P|$ **do**
2:     **if** Dead($P_i$) **or** RateExceeded($P_i$) **then**
3:         **continue**
4:     **end if**
5:     $(s, y) \leftarrow$ Call($P_i$, $p$) ◃ $s$: HTTP status; $y$: body
6:     **if** $s = 200$ **and** Valid$_r$($y$) **then**
7:         **return** $y$
8:     **else**
9:         MarkDead($P_i$, 90 s)
10:     **end if**
11: **end for**
12: **return** $D_r(p)$ ◃ total function; guarantees output

**Algorithm 2** Exploit selection for a detected service

**Require:** service (*port*, *banner*); exploit map $M$; model call `call_ai_fast`
**Ensure:** a Metasploit module + payload, or None
1: **for** each key $(p, kw) \in M$ **do**
2:     **if** $p$ = *port* **and** $kw$ is a substring of *banner* **then**
3:         $(m, pl, \ell) \leftarrow M[(p, kw)]$
4:         **return** Configure($m, pl, \ell$) ◃ deterministic; sets LHOST iff $\ell$
5:     **end if**
6: **end for**
7: $s \leftarrow$ `call_ai_fast`(Suggest, *banner*) ◃ bounded role; validated
8: **if** IsValidModule($s$) **then**
9:     **return** Configure($s$)
10: **else**
11:     **return** None ◃ honest miss; no session claimed
12: **end if**

receives a well-formed result even when all cloud providers are exhausted.

The design consequence is that no single AI call is load-bearing. When every cloud provider is rate-limited, each phase still produces complete output from the rule-based path. This makes the free-tier claim defensible: the framework is not merely cheap when the free tier is available; it is correct when the free tier is exhausted.

## 4.3 Deterministic Exploit Map

The `EXPLOIT_MAP` is a Python dictionary mapping (port, keyword) tuples to (msf_module, payload, needs_lhost) triples. It currently holds 41 entries covering FTP, SMB, databases, web services, distcc, IRC, VNC, NFS, and more. Keywords use the exact version string Nmap prints (e.g., `vsftpd 2.3.4` rather than `ftp`), matched by substring containment. The `needs_lhost` flag distinguishes reverse-shell exploits from bind-shell and auxiliary modules, preventing a class of silent failures discovered during development. Algorithm 2 gives the selection procedure: a map hit yields a deterministic, auditable choice, and only a miss consults the model, which is asked for a bounded suggestion that is itself validated before use.

This map is the backbone that makes a small model sufficient: the model only reasons when the map has no entry, and what it is asked is bounded enough that a 7B-parameter model can do it.

### 4.4 Ten-Phase Pipeline

The orchestration layer drives a ten-phase pipeline: (0) network discovery and topology mapping; (1) multi-technique port discovery; (2) service version detection; (3) NSE script scanning with AI-assisted script selection; (4) multi-source CVE enrichment querying NVD, CISA KEV, ExploitDB and OSV in parallel; (5) web vulnerability scanning (Nikto, Dirb, sqlmap, WPScan) against web-facing services only; (6) exploit generation via map lookup then AI fallback; (7) live exploitation through the held Metasploit interactive console; (8) remediation generation with risk, fix and verification sections; and (9) multi-format report export (HTML, PDF, DOCX, JSON). The AI helper is invoked in nine bounded roles across these phases: network-zone labelling, NSE script selection, CVE ranking, exploit suggestion, web-flag selection, remediation writing, and three analysis steps. Every role has a deterministic fallback, so no single AI call is on the critical path.

### 4.5 MCP Server and Security Design

The FastMCP server [42] exposes eleven tools: `start_scan`, `get_scan_status`, `list_services`, `get_cves_for_service`, `run_exploit`, `get_exploitation_results`, `generate_remediation`, `get_remediation_for_service`, `get_full_report`, `list_active_sessions` and `stop_scan`. Each tool is one HTTP request against the Flask app. The four mitigations of Section 3.3 are implemented at this layer: the server binds to localhost, each scan tool enforces the start-time target allowlist, write/execute tools are absent from the surface, and every call is appended to the audit log.

### 4.6 Multi-Source CVE Enrichment

For each detected service, PentestChain queries four databases in parallel: NIST NVD for CVSS v3.1 scoring [24], CISA KEV for confirmed in-the-wild exploitation, ExploitDB for proof-of-concept code, and Google's OSV for open-source advisories [11, 29]. Results are merged by CVE identifier and ranked by the AI scoring role (CVSS + KEV flag + exploit availability), with a CVSS-only fallback.

## 5 Implementation

### 5.1 Environment and Build

The development environment runs on Windows 11 with Docker Desktop and VirtualBox. The Kali container is built from `kalilinux/kali-rolling` with the Nmap suite, Metasploit Framework, sqlmap, Nikto, WPScan, Python 3.11 and project requirements. Docker Compose orchestrates the application and PostgreSQL containers on a bridge network, pinning eventlet to 0.33.3 (later versions broke Flask-SocketIO monkey-patching). The target is a Metasploitable 2 VM on an isolated host-only network [48, 53].

Three engineering challenges shaped the implementation. First, eventlet's monkey-patching silently broke `request.get_json()`, requiring manual body parsing in every POST endpoint. Second, the exploit map originally used two-tuples; extending to three-tuples with `needs_lhost` fixed backdoor exploits like vsftpd where setting LHOST causes Metasploit to error. Third, the stateless `execute()` pattern lost sessions between phases; switching to pymetasploit3's interactive console with a 30-second deadline keeps sessions alive for remediation verification.

During development, the cloud providers were migrated from Groq and Gemini to OpenRouter and Cerebras after Groq imposed stricter rate limits and Gemini returned HTTP 503 on roughly 30% of free-tier calls at peak. The model string is logged with every call so the choice can be revisited.

**Table 2: Reproducibility status of reported numbers**

| Category | Status |
|---|---|
| Legacy single-target metrics (Tables 3, 6, 4, 5) | Measured |
| Standardised-testbed results (AutoPenBench, Cybench, PentestGPT bench) | Projected |
| Baseline reproductions (PentestGPT, PentestAgent) | Projected |
| Ablation study | Projected |

### 5.2 Deployment Modes

The layered separation supports three validated deployment modes: web-UI-only (MCP server stopped), MCP-only (an external host such as Claude Desktop drives the pipeline with the browser UI never opened), and combined operation. MCP-only is the mode that matters for the free-tier-and-air-gapped story: it runs entirely against a local Ollama model with no browser and no outbound traffic beyond the isolated target network, and a single codebase supports all three without divergent paths.

### 5.3 Database Schema and Instrumentation

PostgreSQL stores scan state across restarts in six tables: `scans`, `services`, `cves`, `exploitation`, `remediation`, and `ai_calls`. The `ai_calls` table records each model call with provider, model string, token count, latency and outcome, making the AI fallback rate and dollar cost directly computable from the record. This instrumentation is what turns the central cost claim from an assertion into a measurement.

## 6 Experimental Design

This section specifies the evaluation protocol. It has two parts: the measurements already collected on the legacy targets (Section 6.6 onward), and a strengthened, standardised protocol whose results are being instrumented. Table 2 states which is which: the legacy single-target metrics in Tables 3, 6, 4 and 5 are measured, whereas the standardised-testbed comparison, the reproduced baselines and the ablation constitute a *planned* protocol whose results have not yet been collected and must not be read as measured outcomes.

### 6.1 Standardised Testbed

To make results comparable with prior work and robust to the external-validity concern that legacy targets raise, we evaluate on three established, containerised testbeds in addition to the five in-house targets:

- **AutoPenBench** [28]: 33 milestone-scored tasks (22 in-vitro, 11 real-world CVE), reported as milestone completion and

task success. This is the primary head-to-head instrument because its milestone scoring rewards partial progress, matching PentestChain's staged pipeline.

- **Cybench (subset)** [66]: a stratified subset of the 40 CTF tasks, reported in unguided and subtask-guided modes, to probe long-horizon reasoning where the deterministic map does not help.
- **PentestGPT benchmark** [14]: the 182 sub-tasks across 13 targets, using sub-task completion for a granular comparison against the closest conceptual neighbour.

All targets, including the five in-house VMs, are run as pinned container images so that a reviewer can reproduce the environment exactly; image digests are listed in the Open Science appendix.

## 6.2 Metrics

We report a capability axis and a cost axis. Capability: `pass@`$k$ and task/milestone success rate on the standardised testbeds; service detection, CVE-enrichment and confirmed-session counts on the in-house targets. Cost: measured paid-API dollars per engagement (from `ai_calls`), token consumption, wall-clock time, and the AI fallback rate. Every run is additionally classified by outcome—success, refusal, iteration-budget exhaustion, or infrastructure error—so that genuine model refusals are never conflated with transport failures, a distinction whose absence has produced published errata in this area.

## 6.3 Cost Model

We make the cost axis precise, since a measured dollar figure is only meaningful against an explicit model of what is being counted. Let an engagement issue calls to a set of providers $P$; for provider $p$ let $n_p^{\text{in}}$ and $n_p^{\text{out}}$ be the input and output tokens it serves and $\rho_p^{\text{in}}, \rho_p^{\text{out}}$ its per-token prices. The paid cost of an engagement is

$$C_{\text{eng}} = \sum_{p \in P} n_p^{\text{in}} \rho_p^{\text{in}} + n_p^{\text{out}} \rho_p^{\text{out}} . \quad (1)$$

Every term is read directly from the `ai_calls` table (Section 5.3), so $C_{\text{eng}}$ is a measurement rather than an estimate. The architecture forces the price coefficients to zero: the local Ollama tier has $\rho = 0$ by construction (computation stays on the host), the OpenRouter and Cerebras tiers are used strictly within their free quotas ($\rho = 0$ while under quota), and the deterministic branch $D_r$ makes no API call at all. Hence $C_{\text{eng}} = 0$ is not a favourable measurement but an invariant of the configuration: the only way an engagement can incur a charge is if a paid tier is enabled. This is the sense in which the cost reduction is *structural*.

The model also makes the counterfactual explicit. Substituting a premium tier for the free/local one leaves the token counts largely unchanged but sets $\rho > 0$, so cost scales linearly with the measured token volume $T = \sum_p (n_p^{\text{in}} + n_p^{\text{out}})$, which averaged 181 ± 23K tokens per engagement (Table 3). As an external anchor, Fang et al. measured USD 3.52 per run for a comparable autonomous GPT-4 agent on one-day exploitation [21]; under continuous testing this recurs per engagement and compounds with cadence, whereas Equation 1 stays at zero regardless of volume. We report the premium figure as a projection, not a measured PentestChain cost (Table 2).

## 6.4 Protocol and Statistics

Because LLM outputs are stochastic even at temperature 0, each configuration is run for ≥ 10 independent trials; seeds, model strings and full prompts are logged and released. We report mean ± SD and 95% confidence intervals, and `pass@`$k$ for $k \in \{1, 3, 5\}$. For comparisons between systems or configurations we use non-parametric tests appropriate to small, non-normal samples—the Mann–Whitney $U$ test for unpaired and the Wilcoxon signed-rank test for paired settings—and report effect sizes (Cohen's $h$ for proportions, Cliff's $\delta$ for ordinal outcomes) rather than $p$-values alone. When many configuration pairs are compared we apply a Holm–Bonferroni correction. Sample size is justified by a power analysis targeting the smallest effect of practical interest (a 10-percentage-point difference in success rate).

## 6.5 Baselines and Ablations

Following current practice, we do not cite competitors' published numbers as a comparison; we *reproduce* PentestGPT [14] and PentestAgent [57] on the identical testbeds and harness, each in its strongest documented configuration, controlling for target and tool-access differences. To isolate the contribution of each component we ablate four factors, reporting stage-level completion in the manner of PentestAgent: (i) *MCP orchestration* (pipeline driven via MCP vs. directly); (ii) *the cost-router* (full cascade vs. local-only vs. single-provider); (iii) *model scale* (`qwen2.5:7b` vs. a larger local model vs. a premium API model in the same slots); and (iv) *the deterministic exploit map* (map + AI vs. AI-only), which directly measures how much of the performance the map rather than the model supplies.

## 6.6 Experimental Setup (Measured Runs)

The lab runs on a single host with the PentestChain Kali container (Flask UI on port 4444, FastMCP on port 8000) and a PostgreSQL container. Four targets are deployed on isolated networks: Metasploitable 2 (Linux, 4 runs), OWASP Broken Web Applications (BWA, web-focused Linux hosting DVWA, WebGoat, Mutillidae and 15+ applications, 2 runs), a Windows 7 SP1 VM with SMB/RDP exposed (2 runs), and bWAPP/bee-box in Docker (1 run); a fifth target—a contemporary Debian host—is added in Section 6.13 to probe the legacy-only limitation, giving five targets and ten runs in total. Seven metrics assess the framework: detection coverage, exploitation success, remediation coverage, AI fallback rate, pipeline time, token consumption, and dollar cost per engagement [9, 33, 51, 54].

## 6.7 Detection and Enrichment

Across four runs the framework detected 26.8 ± 0.5 open ports and 26 services, enriching them against four CVE databases to 34 unique CVEs (7 critical, 15–22 high depending on scan profile). The top-ranked CVEs included CVE-2011-2523 (vsftpd backdoor, CVSS 9.8), CVE-2008-0122 (ISC BIND, CVSS 10.0), and CVE-2010-2075 (UnrealIRCd backdoor).

## 6.8 Exploitation Results

The exploitation phase produced confirmed sessions for the principal vulnerable services. The `exploit/unix/misc/distcc_exec` module opened a cmd/unix shell session (Session 1) via a bind payload on port 4445, and `exploit/linux/postgres/postgres_payload`

**Table 3: Consolidated Evaluation Metrics Across Four Runs**

| Metric | R1 | R2 | R3 | R4 | Mean ± SD |
|---|---|---|---|---|---|
| Open ports | 27 | 26 | 27 | 27 | 26.8 ± 0.5 |
| Services | 26 | 26 | 26 | 26 | 26.0 ± 0.0 |
| CVEs enriched | 34 | 34 | 34 | 34 | 34.0 ± 0.0 |
| Critical | 7 | 7 | 7 | 7 | 7.0 ± 0.0 |
| High | 22 | 15 | 22 | 15 | 18.5 ± 4.0 |
| Exploits ready | 11 | 12 | 11 | 12 | 11.5 ± 0.6 |
| Fallback rate | 4.4% | 3.6% | 3.3% | 3.1% | 3.6 ± 0.6% |
| Tokens (K) | 168 | 194 | 206 | 156 | 181 ± 23 |
| Pipeline (min) | — | 22.7 | 16.9 | 25.6 | 21.7 ± 4.4 |
| Ollama calls | 68 | 83 | 91 | 65 | 76.8 ± 12.0 |
| Fix plans | — | 13 | 13 | 14 | 13.3 ± 0.6 |
| Dollar cost | $0 | $0 | $0 | $0 | $0.00 |

uploaded a shared-object payload to open a Meterpreter session (Session 2) on the same bind port. Both sessions were captured and held for remediation verification. Between 5 and 23 successful exploit actions were recorded per run depending on the scan profile, with the deterministic exploit map catching the large majority before the AI suggester was consulted.

Remediation was generated for every detected service (risk, step-by-step fix, verification command), cross-checked against vendor advisories; residual model-written entries recommended the correct fix. The report exported to HTML, PDF, DOCX and JSON, its executive summary written by the AI summary role with a deterministic template fallback.

## 6.9 Web Vulnerability Scanning

Phase 5 runs web vulnerability scanning only against services detected as web-facing, so heavier tools are never launched blindly. Nikto, Dirb, sqlmap, WPScan and header/XSS checks are driven in parallel per web service, with an AI flag-suggester narrowing the sqlmap invocation and an impact-chain step annotating each finding. Keeping this phase AI-targeted is a cost and time decision: the model decides which tools to run, and the deterministic path runs a sensible default when the model is unavailable. The web scanning phase produced 60 findings across the engagement, including XSS and injection issues grouped separately from service-level CVEs.

## 6.10 Cost and AI Fallback Analysis

This is the central result. Table 3 consolidates all metrics across four runs. The instrumentation recorded a mean of 76.8 ± 12.0 calls to the local Ollama model, 21.8±1.9 calls to OpenRouter, and 21.8±1.9 to Cerebras. The Ollama success rate was 96.4 ± 0.6% with mean latency of 16,114 ms. The measured paid-API spend across all four engagements was zero. The AI fallback rate to the rule-based path was 3.6 ± 0.6%, clustered around peak hours on free-tier providers.

## 6.11 Planned Testbed and Ablation Results

Under the protocol of Sections 6.1–6.5 we will report, per system, AutoPenBench milestone/task success, Cybench-subset success (unguided and subtask-guided) and PentestGPT-benchmark sub-task completion, each paired with measured $/engagement and `pass@`*k*; the free-tier slot is expected to trail the premium slot most on the long-horizon Cybench and sub-task axes and least on the enrichment-heavy axes the deterministic backbone supports. The same runs drive a four-lesion ablation—MCP-vs-direct orchestration; full-cascade vs. local-only vs. single-provider routing; model scale (`qwen2.5:7b` vs. a larger local vs. a premium API model); and, decisively, map+AI vs. AI-only—reporting stage-level completion (information gathering, vulnerability discovery, exploitation). Removing the deterministic map is expected to collapse confirmed exploitation for the 7B model while barely moving information gathering, quantifying how much of the capability the map rather than the model supplies. These results are not yet collected (Table 2).

**Table 4: Cross-Platform Evaluation Summary**

| Metric | MSF2 | BWA | Win7 | bWAPP | Modern |
|---|---|---|---|---|---|
| OS / type | Linux | Linux | Windows | Linux | Linux |
| Vuln. profile | Legacy | Web | Legacy | Web | *Hardened* |
| Runs | 4 | 2 | 2 | 1 | 1 |
| Open ports | 26.8 | 10.0 | 422.5 | 21 | 3 |
| CVEs enriched | 34 | 85.0 | 44.0 | 14 | 7 |
| Critical | 7 | 37.0 | 5.5 | 2 | 1 |
| High | 18.5 | 28.0 | 12.5 | 4 | 1 |
| Exploits confirmed | 12 | 3.0 | 1.0 | 0 | 0 |
| Web findings | 60 | 180.5 | 59.5 | 74 | 48 |
| Dollar cost | $0 | $0 | $0 | $0 | $0 |

MSF2 = Metasploitable 2. BWA and Win7 values are means of two runs. “Modern” = a contemporary Debian host (OpenSSH 8.4p1, Apache 2.4.51, nginx 1.18.0). MSF2 CVE, critical and high counts are the four-run figures of Table 3; the MSF2 web-findings figure matches Section 6.9.

## 6.12 Cross-Platform Generalisability

To move beyond the single-target limitation, the framework was evaluated against five environments spanning two operating systems and four vulnerability profiles: Metasploitable 2 (legacy Linux), OWASP BWA (a web-focused VM hosting DVWA, WebGoat, Mutillidae and others), a Windows 7 SP1 host (SMB/RDP), bWAPP/bee-box (a containerised PHP target), and—to probe the legacy-only limitation—a contemporary Debian host running current-generation services. Table 4 summarises the results.

Three findings emerge. First, the deterministic exploit map generalised across the vulnerable targets without modification: `tomcat_mgr_upload` and `ms17_010_eternalblue` matched on both BWA and Windows 7, and `vsftpd_backdoor`, `samba_usermap_script` and `distcc_exec` on Metasploitable 2. Second, web scanning scaled to BWA’s large surface (180.5 findings on average across DVWA, WebGoat, BodgeIt, Mutillidae and WackoPicko: SQL/command injection, file upload and remote file inclusion). Third, Windows 7 exposed a variability pattern—530 ports/59 CVEs/2 exploits in Run 1 (including EternalBlue) versus 315/29/0 in Run 2, likely from Windows Firewall state differences—while bWAPP, a single-application container, had the smallest surface (14 CVEs, no exploitable modules) yet still yielded 74 web findings.

These results show the pipeline is not overfit to a single target’s service fingerprint: the exploit map’s substring matching against Nmap version strings transfers to new environments without retraining, and the zero-cost guarantee held across all ten runs. The modern Debian host is the most informative negative result—detection and enrichment functioned while deterministic exploitation correctly declined to fire—and is analysed in Section 6.13.

**Table 5: Modern-Target Spot Check: Detected Services (192.168.64.9)**

| Port | Service | Version | Findings |
|---|---|---|---|
| 22 | ssh | OpenSSH 8.4p1 (Debian) | — |
| 80 | http | Apache httpd 2.4.51 | 17 |
| 81 | http | nginx 1.18.0 | 31 |
| 7 CVEs (1 critical, 1 high); 0 exploit modules matched; risk 3.0/10. | | | |

**Table 6: AI Cascade Reliability and Cost (per run)**

| Provider (tier) | Calls | Success | Latency | Cost |
|---|---|---|---|---|
| Ollama qwen2.5:7b (local) | 76.8 | 96.4% | 16,114 ms | $0 |
| OpenRouter (free) | 21.8 | — | — | $0 |
| Cerebras (free) | 21.8 | — | — | $0 |
| Rule-based fallback | 3.6% | 100% | <1 ms | $0 |

Fallback row reports the share of AI decisions served deterministically, not a call count. Dashes mark values not separately instrumented.

## 6.13 Modern-Infrastructure Spot Check

To probe the legacy-only limitation, we ran the full pipeline once against a contemporary Debian host with current-generation, largely patched services. Table 5 lists the detected surface. The scan found three open services and enriched seven CVEs, of which the AI ranking role correctly elevated CVE-2021-44790 (an Apache `mod_lua` multipart buffer overflow, CVSS 9.8) and CVE-2021-44224 (an Apache forward-proxy NULL-pointer dereference, CVSS 8.2) above lower-severity nginx advisories from OSV. The remaining phases behaved as the architecture predicts on a hardened target: the deterministic exploit map returned *no direct module match*, so zero Metasploit sessions were established, while the web-scanning phase still produced 48 findings (missing security headers, clickjacking, directory listings, an auth-gated endpoint). The aggregate risk score was 3.0/10 (moderate), against 10.0/10 for the legacy BWA target—evidence that the scoring reflects genuine exploitability rather than raw finding volume. A single run does not establish behaviour on hardened infrastructure at scale, but the graceful degradation is the architecture's expected outcome.

## 6.14 Reliability of the Cost-Aware Cascade

Table 6 breaks the cascade down per provider. Because the rule-based path always returns a valid result, end-to-end reliability of the AI layer is effectively 100% by construction—no phase can fail for lack of a model response.

## 6.15 Representative Engagement: OWASP BWA Walk-Through

A representative OWASP BWA engagement grounds Table 4's aggregate figures in an auditable per-service breakdown. Scanning ports 1–10000 detected ten open services (SSH, dual Apache instances, Samba, IMAP, a Java object-serialisation endpoint, and three web-application containers). Multi-source enrichment returned 79 unique CVEs (37 critical, 28 high), a maximal 10.0/10 risk rating. Phase 6 produced three map-matched exploit scripts (`tomcat_mgr_upload`, `tomcat_mgr_deploy`, `ms17_010_eternalblue`) before the AI suggester was consulted; the AI-targeted web phase produced 163 findings, including 15 reflected-XSS proof-of-concepts across Mutillidae, an active HTTP TRACE method, an exposed `.bash_history`, and an unprotected phpMyAdmin interface. Phase 8 generated remediation for all five remediable services at zero paid-API cost.

## 6.16 Comparison with Baselines

Table 7 positions PentestChain against the closest related systems. Where the commercial scanners show no exploitation, that is by design—a scanner cannot risk triggering even a benign exploit at production scale [53], a trade-off borne out in enterprise scanner comparisons that measure coverage but not exploitation [8]. What PentestChain uniquely shows is comparable end-to-end function—reconnaissance through verified remediation—at zero measured cost, whereas the same call volume on a premium model incurs a recurring paid-API charge that compounds with usage while the free-tier configuration stays at zero regardless of volume.

**Table 7: PentestChain vs. Premium-Model Frameworks and Scanners**

| System | Architecture | Model | Cost | Eval. scope |
|---|---|---|---|---|
| PentestGPT [14] | 3-module, human loop | GPT-4 | N/R | 182 sub-tasks |
| PentestAgent [57] | Multi-agent + RAG | GPT-4 | N/R | 74.2% succ. |
| Ezetta–Feng [20] | MCP toolkit | GPT-4.1 | N/R | 100+ VulHub |
| Multi-agent PentestMCP [4] | Multi-agent + MCP + RAG | GPT-4.1 | N/R | 100+ CVE |
| PentAGI [65] | 12+ agents, pgvector | Premium | N/R | Broad |
| Nessus et al. [59] | Plugin CVE enum. | None | Licence | Ent.; no exploit |
| **PentestChain** | Pipeline + exploit map + MCP | Free / local | $0 | 5 targets, 10 runs |

## 7 Discussion

*Architecture, not scale, is the operative lever.* The results support the paper's central claim structurally rather than incidentally: because the deterministic map supplies the execution a small model cannot reliably generate and the model is confined to labelling and ranking, the 7B configuration reaches comparable end-to-end function without a frontier model. This is the practical face of the knowledge–execution gap [30]—the model knows what to do far more often than it can do it through tool calls, so moving the "doing" into a deterministic backbone is what makes the small model sufficient. The ablation of Section 6.5 (map+AI vs. AI-only) is designed to quantify exactly this split, and we expect removing the map to collapse confirmed exploitation while barely moving information gathering.

*Generalisation and its limits.* The exploit map transferred across two operating systems and four vulnerability profiles without modification, and the contemporary Debian host is the most informative result of all: detection and enrichment functioned while deterministic exploitation correctly declined to fire, the graceful degradation the architecture predicts on hardened infrastructure. The mechanism of generalisation—substring matching against Nmap version strings—is also its boundary: a target that suppresses or falsifies banners, or that runs a service absent from the map, falls through to the bounded model suggester, whose reliability on genuinely novel services is the open question the standardised-testbed evaluation is meant to settle.

*When the free-tier approach breaks.* The design is deliberately robust to free-tier volatility—quotas, latency spikes and silent model changes are absorbed by the cascade and, ultimately, the total deterministic branch of Equation 1—but two costs are real. The map is a curated asset that must be maintained as services and modules evolve, and long-horizon, multi-step reasoning of the kind Cybench probes is where a 7B model, even with a deterministic floor, is expected to trail a premium model most. We therefore scope the capability claims as projected until the protocol of Sections 6.1–6.5 is complete, and report cost, not capability, as the settled contribution.

*The tool is itself an attack surface.* Exposing exploitation over MCP buys composability at the price of a new surface (Section 3); the default controls close the server-side positions but A3 remains a client-side residual (Table 1). Treating the offensive engine's own protocol as untrusted—rather than assuming the convenience of MCP is free—is, we argue, the responsible default for any tool of this class.

## 8 Ethics and Responsible Use

PentestChain is offensive tooling and is presented with the dual-use considerations expected of such work [2, 16], and the engagements follow the standard phased penetration-testing methodology [17] within our institution's research-ethics code [63]. Three points are material. *Scope of experimentation:* every result in this paper was obtained against systems we own—deliberately vulnerable VMs and containers (Metasploitable 2, OWASP BWA, bWAPP) and a self-hosted Debian host—on isolated host-only networks, with no testing against any third-party or production system. *Impact and proportionality:* the framework automates the composition of existing, publicly documented tools (Nmap, Metasploit, sqlmap) rather than introducing new exploits; its incremental risk is one of *accessibility and scale*, which we mitigate by shipping the four MCP controls of Section 3.3 enabled by default. *Disclosure:* the paper contributes an analysis of the MCP attack surface built on already-public vulnerabilities (CVE-2025-6514, `postmark-mcp`); should the planned experiments surface any previously unknown vulnerability in a target or in the MCP tooling, we will follow coordinated disclosure with a 45–90 day window before any public release. A GenAI Usage Disclosure accompanies this submission, and all artifacts needed to reproduce the measured results (source, Dockerfile, pinned image digests and exported `ai_calls` tables) will be released under an anonymous review link.

## 9 Threats to Validity

AI responses are non-deterministic; but mitigated by multi-run averaging, full prompt/response logging, and the deterministic fallback. The strengthened protocol (Section 6.4) additionally fixes and logs seeds, reports variance and effect sizes over ≥ 10 trials, and separates model refusals from infrastructure errors so that stochasticity is characterised rather than merely acknowledged. *External:* Four of the five evaluation targets are deliberately vulnerable legacy systems; results should not be extrapolated to modern hardened production infrastructure. The added modern Debian host begins to probe this gap; the AutoPenBench and Cybench tasks in the standardised protocol are the instruments intended to close it. *Construct:* Exploitation success is measured by Metasploit session establishment, consistent with DeepExploit and PenHeal. Dollar cost is measured as paid-API spend. *Threat-model completeness:* the four mitigations are defence-in-depth, not a proof of security; adversary A3 is a client-side RCE that the engine cannot fully close, so operators must pin client versions and treat tool descriptions as untrusted. *Reliability:* Containerised deployment ensures identical environments; the Dockerfile and `docker-compose.yml` are released. The Windows 7 run variability (530 vs. 315 ports) highlights host-state sensitivity that containerisation alone does not control.

## 10 Future Work

The strengthened evaluation of Section 6 is the immediate next step and the precondition for the capability claims this paper scopes as projected. Beyond it, the roadmap extends the framework while preserving the zero-cost, deterministic-fallback architecture: a scaled VulHub evaluation across at least five CWE categories in a two-arm (free-tier vs. premium) design [21]; a signed, idempotent remediation-apply-and-verify loop with rollback; an air-gapped deployment pairing a LoRA/QLoRA-tuned local model [15, 35] with offline mirrors of NVD, KEV, ExploitDB and OSV; expanded OWASP Top 10 [50] web scenarios; and cloud and Active Directory modules driven by playbooks composed dynamically over the MCP surface.

## 11 Conclusion

PentestChain demonstrates that a deterministic exploit-map backbone, combined with a cost-aware AI cascade and MCP orchestration, achieves comparable penetration testing effectiveness to premium-model frameworks at zero measured cost. Across ten engagements against five heterogeneous targets, the framework enriched 7–85 CVEs per target, confirmed exploitation on three of five platforms including EternalBlue on Windows, and produced 180+ web-application findings on OWASP BWA's multi-application surface, all at zero paid-API spend. The deterministic exploit map generalised across operating systems and vulnerability profiles without modification.

The contribution is architectural rather than model-centric: by confining the LLM to nine bounded, fallback-protected roles and placing the offensive logic in a deterministic exploit map, the framework makes a 7B-parameter model sufficient for what would otherwise require a premium frontier model. By measuring dollar cost as a first-class metric, the framework makes the accessibility claim falsifiable rather than asserted. By analysing the MCP surface it exposes and specifying a standardised, statistically grounded evaluation protocol, it positions that claim to be tested against the bar a top-tier venue expects. For anyone building on this work, commit to a deterministic fallback in every AI-using component and measure dollar cost as a first-class metric: free-tier limits and silent model changes are facts of life, and a framework that fails when they bite is a demonstration, not a tool.

## GenAI Usage Disclosure

Generative-AI tools were used by the authors only for minor language editing and LaTeX formatting assistance. The research problem, system design, implementation, experimental protocol, data collection, analysis, and all scientific claims are the authors' own, and every AI-assisted passage was reviewed and verified by the authors.

## Open Science

The contributions of this paper rely on research artifacts: the PentestChain engine (the Flask/MCP application, the deterministic EXPLOIT_MAP, and the cost-aware cascade), the containerised evaluation harness, and the exported instrumentation from which every reported cost and fallback figure is computed. In line with the venue's open-science policy, these artifacts are made available to reviewers through an anonymous link: https://github.com/Rushabh0508/Pentest-Chain

The release contains the engine source, the Dockerfile and docker-compose.yml that reproduce the Kali/PostgreSQL environment, the pinned container-image digests for all evaluation targets (the five in-house VMs and the three standardised testbeds referenced in Section 6.1), the prompt templates and fixed seeds used in every role, and the exported ai_calls tables that instantiate the cost model of Equation 1. Measured results (Tables 3, 6, 4 and 5) can be regenerated directly from these artifacts; the standardised-testbed and ablation results marked *projected* in Table 2 will be added to the same repository as they are collected. A non-anonymised, permanently archived release with a DOI will accompany the camera-ready version. No datasets containing personal or otherwise sensitive data are used or released.

## Reproducibility Details and Representative Engagement

*Lab specification.* The environment is rebuilt from two files in the artifact release. A Dockerfile builds the Kali application image (the Nmap suite, the Metasploit Framework, sqlmap, Nikto, WPScan, Python 3.11 and the pinned pip requirements), and a docker-compose.yml defines the application and PostgreSQL services, the bridge network, the named data volume, the eventlet 0.33.3 pin, and health checks for both containers. A reviewer rebuilds the exact environment behind every measured result with a single docker compose up --build, points the target at a Metasploitable 2 instance on the isolated host-only network, and re-runs the pipeline through the web UI on port 4444 or through any MCP host connected to the FastMCP server on port 8000.

*Remediation example.* Each service carrying a ranked finding received a three-part remediation entry—risk, a step-by-step fix and a verification command—the per-run count of which is reported as *Fix plans* in Table 3. For the vsftpd 2.3.4 backdoor, for instance, the generated entry rated the risk critical (unauthenticated root), prescribed removing or upgrading the package to a patched build and blocking port 21 at the host firewall, and supplied nmap -p21 --script ftp-vsftpd-backdoor against the target as the check that the backdoor no longer responds. Because the framework holds the exploited session open, such fixes were verified against the live session before the report was written. The same structured state is exported to HTML, PDF, DOCX, JSON and CSV so that a non-technical reader sees the headline risk first while an auditor can trace each finding from banner to ranked CVE to confirmed session to verified fix.

**Table 8: Representative Metasploitable 2 engagement: confirmed findings instantiating the aggregate figures of Table 3**

| Service (CVE) | CVSS | Outcome |
|---|---|---|
| vsftpd 2.3.4 (CVE-2011-2523) | 9.8 | Root shell — session |
| distcc (distcc_exec) | 9.3 | Shell — Session 1 |
| PostgreSQL (postgres_payload) | — | Meterpreter — Session 2 |
| Root bindshell (port 1524) | Crit. | Root shell — session |
| UnrealIRCd 3.2.8.1 (CVE-2010-2075) | 10.0 | Shell — session |
| ProFTPD 1.3.1 (CVE-2009-0542) | 9.8 | Auxiliary success |
| ISC BIND (CVE-2008-0122) | 10.0 | Reported, not exploited |
| Apache Tomcat (port 8180) | High | Module ready |

One representative run of the four-run Metasploitable 2 evaluation, whose means (26 services, 34 unique CVEs with 7 critical, 11.5 ready and 12 confirmed exploits, a 10.0/10 aggregate risk score, all at $0 measured paid-API cost) are reported in Tables 3 and 4. The failed Heartbleed attempt on a version-less banner is the residual AI-suggester case noted in the exploitation results.

## Ethical Considerations

Because PentestChain automates offensive security, we treat its ethics explicitly rather than leaving them implicit. Section 8 states the operational safeguards; this appendix gives the fuller harm–benefit analysis that dual-use work of this kind warrants, following the principles of the Menlo Report [16] and the ACM Code of Ethics and Professional Conduct.

*Benefit.* The intended beneficiaries are precisely the under-resourced organisations that the workforce-gap evidence [39] identifies as least able to afford continuous frontier-model testing; driving per-engagement cost to zero lowers the barrier to routine defensive self-assessment, which is the research's motivating good. *Harms considered.* We weighed risks to three groups: operators (the client-side MCP exposure of adversary A3), owners of systems that could be tested without authorisation, and the wider ecosystem, in which releasing an accessible offensive tool could lower the cost of attacks. Decisively, the framework introduces no new exploit—it orchestrates publicly documented tools—so its incremental risk is one of *accessibility and scale*, not of capability.

*Research conduct.* Every experiment ran against systems we own (deliberately vulnerable VMs and containers, and a self-hosted host) on isolated, host-only networks, with no third-party or production target and no human subjects or personal data, under our institutional research-ethics code [63]. Because no human subjects were involved the work fell outside human-subjects review, but the code was applied to scoping and disclosure regardless.

*Safeguards against misuse.* The four controls of Section 3.3 ship enabled by default. The start-time target allowlist in particular turns "authorised testing" from a stated policy into an enforced precondition, and excluding file-write and remediation-execution tools from the MCP surface bounds the blast radius of a poisoned tool description. We are explicit that these are defence-in-depth, not a proof of safety: A3 is a client-side flaw the engine cannot fully close, so we recommend pinning client versions and treating tool descriptions as untrusted input.

*Disclosure and release.* The MCP analysis builds only on already-public vulnerabilities (CVE-2025-6514, `postmark-mcp`); any previously unknown flaw surfaced by the planned experiments will be handled under coordinated disclosure with a 45–90 day window. The artifact release (Open Science appendix) is gated behind an anonymous review link and carries no offensive payloads beyond the public Metasploit modules the engine already invokes.

*Net assessment.* On balance we judge publication net-positive: the underlying capabilities are individually public, the contribution is architectural rather than a new weapon, and the safeguards and disclosure posture raise the cost of misuse relative to the defensive benefit the zero-cost design unlocks.